\documentclass[12pt,preprint]{aastex}

\begin{document}

\title{Observations of Intrahour Variable Quasars: Scattering in our Galactic Neighbourhood\footnote{This is a preprint of an article whose final and definitive form has been accepted for publication in Astronomical and Astrophysical Transactions, copyright 2007 Taylor and Francis.}}

\author{H.~E.~Bignall}
\affil{Joint Institute for VLBI in Europe, Postbus 2, 7990AA Dwingeloo, the Netherlands}
\email{bignall@jive.nl}

\author{D.~L.~Jauncey, J.~E.~J.~Lovell and A.~K.~Tzioumis}
\affil{Australia Telescope National Facility, PO Box 76, Epping, NSW 1710, Australia} 

\author{J-P.~Macquart}
\affil{NRAO Jansky Fellow based at Department of Astronomy, Mail Code 105-24 Robinson, Caltech, Pasadena CA 91125 U.S.A.}

\author{L.~Kedziora-Chudczer}
\affil{Institute of Astronomy, School of Physics A28, University of Sydney, NSW 2006, Australia}

\begin{abstract}
Interstellar scintillation (ISS) has been established as the cause of
the random variations seen at centimetre wavelengths in many compact
radio sources on timescales of a day or less. Observations of ISS can
be used to probe structure both in the ionized insterstellar medium of
the Galaxy, and in the extragalactic sources themselves, down to
$\mu$as scales.  A few quasars have been found to show large
amplitude scintillations on unusually rapid, intrahour timescales.
This has been shown to be due to weak scattering in very local
Galactic ``screens'', within a few tens of parsec of the Sun.  The
short variability timescales allow detailed study of the scintillation
properties in relatively short observing periods with compact
interferometric arrays.   The three best-studied ``intrahour
variable'' quasars, PKS~0405$-$385, J1819+3845 and PKS~1257$-$326,
have been instrumental in establishing ISS as the principal cause of
intraday variability at centimetre wavelengths.  Here we review the
relevant results from observations of these three sources.
\end{abstract}

\section{Introduction}

Intraday variability (IDV) was discovered at centimetre wavelengths in
the mid-1980s \citep{wit86,hee87}. For over a decade there was much
debate over whether the variability could be source-intrinsic, or
whether it was caused by extrinsic mechanisms, namely gravitational
microlensing or interstellar scintillation (ISS). Intrinsic
explanations imply very high brightness temperatures, requiring
Doppler factors of 50--200 for consistency with the Inverse Compton
limited brightness temperature of $\sim 10^{12}$\,K. Over the last
decade much observational evidence has accumulated to support ISS as
the principal mechanism for centimetre wavelength IDV.

The largest survey for IDV to date is the MASIV 5 GHz VLA Survey of
more than 500 compact, flat-spectrum radio sources \citep{lov2003}.
The MASIV Survey revealed variability on timescales of up to 3 days
(the duration of the MASIV observing sessions) with typical modulation
indices $\sim 1-10$\%, in more than half of the observed sources
during one or more epochs. Among other factors, the Galactic latitude
distribution of the scintillating sources provides strong evidence of
an interstellar origin of the observed variability.  While ISS of
flat-spectrum radio sources is common, variability on timescales of a
few hours or less and with rms amplitude modulation more than $\sim
10$\% is extremely rare.  At the extreme end of the IDV spectrum,
three quasars are known to show large variations on timescales of less
than 1 hour. These are PKS~0405$-$385 \citep{ked97}, J1819+3845
\citep{dtdb2000} and PKS~1257$-$326 \citep{big2003}. Because of the
short timescales of the fluctuations, they can be well sampled in a
typical 12 hour observing session with an interferometer such as the
Australia Telescope Compact Array (ATCA) or Westerbork Synthesis Radio
Telescope (WSRT), and this has enabled detailed studies of the
variability characteristics. As described by \citet{mj2002}, ISS can
be used to probe both the radio structure of the high-brightness
source components and the insterstellar medium (ISM) responsible for the
rapid variations. In this paper we review the discovery, observations
and analysis of the three well-studied intrahour variable (IHV)
sources.

\section{The Intrahour Variable Quasars}

\subsection{PKS~0405$-$385}
\begin{figure}[h]
\begin{center}
  \includegraphics[width=15cm]{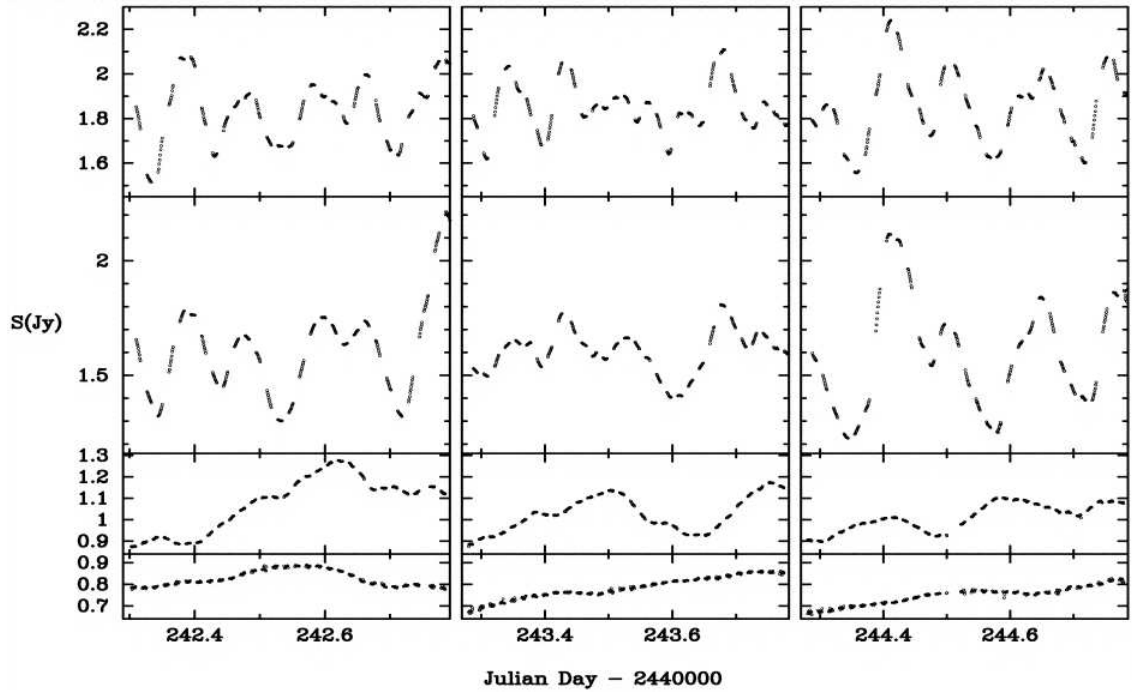}
  \caption{Rapid variations observed in PKS~0405$-$385 over three days
    in June 1996, from \citet{ked97}. The observed frequencies are,
    from top to bottom, 8.6, 4.8, 2.4 and 1.4\,GHz.}\label{fig-0405}
\end{center}
\end{figure}

The discovery of hourly variations in the southern $z = 1.285$ quasar
PKS 0405$-$385 at 4.8 and 8.6\,GHz caused serious difficulties to
explain the variability as intrinsic to the source \citep{ked97}.  As
shown in Fig.~\ref{fig-0405}, the variability was so large and rapid,
with changes of up to 50\%, or $\sim 1$\,Jy, in an hour or less at
5\,GHz, that the implied brightness temperature for intrinsic
variability was in excess of $10^{21}$\,K and so the authors were led
to consider ISS. The observed frequency dependence of the modulation
amplitude and timescales were both consistent with weak scattering at
frequencies of 5\,GHz and above, and strong scattering at frequencies
below 5\,GHz, with the largest amplitude variations close to the
transition frequency \citep{wal98}. Remarkably, the dramatic
variations ceased after a few months.

Confirming evidence for an ISS origin of the rapid variability in
PKS~0405$-$385 came with the next episode of IDV in late 1998. The
variations were sufficiently rapid and strong that it was feasible to
time the variability patterns at two widely spaced telescopes and
search for any time delay between the patterns. Such an experiment can
only be done if the variability timescale is sufficiently short,
typically an hour or less, and the measurement accuracy high enough to
detect flux density changes in tens of seconds, as was the case with
PKS~0405$-$385.  A successful experiment was undertaken between the
ATCA and the Very Large Array (VLA) at 4.8 and 8.6\,GHz. At the
southerly declination of $-39^{\circ}$ the geometry is less than
ideal, with very little overlap between PKS~0405$-$385 rising at the
ATCA and setting at the VLA,  but nevertheless a significant time
delay of $140 \pm 25$ seconds was found, with the pattern arriving
first at the VLA \citep{vsop:jau2000}.  Such a time delay demonstrates
unequivocally that ISS is the mechanism responsible for the dramatic
variability in this source, at the same time ruling out intrinsic
variability. The time delay also constrains the velocity at which the
scintillation pattern drifts across the baseline. 

Unfortunately once again the rapid variability in PKS~0405$-$385
ceased before the measurements could be repeated, and did not reappear
again during the course of an ATCA monitoring program which lasted
until mid-2002.  In 2004, however, PKS~0405$-$385 was once again found
to be showing large and rapid variations \citep{cim2004}. ATCA
observations in early 2006 revealed extremely rapid fluctuations on
timescales much less than 1 hour, allowing the pattern time delay
between the VLA and the ATCA to be measured again, this time to very
high accuracy, $177.2 \pm 4.5$ seconds (Kedziora-Chudczer, presented
at ``Challenges of Relativistic Jets'' meeting in Cracow, Poland, June
2006).

\begin{figure}[h]
\begin{center}
  \includegraphics[width=13cm]{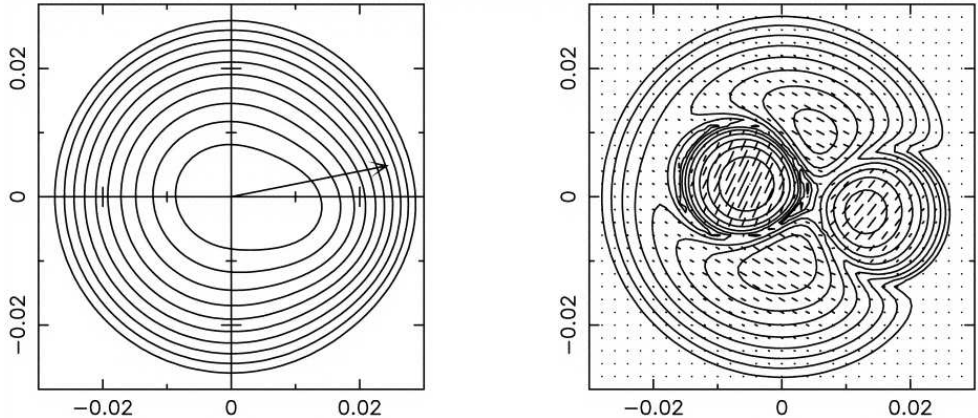}
  \caption{A model for the $\mu$as-scale polarized structure of
    PKS~0405$-$385, from \citet{ric2002}. The angular scale is shown
    in milliarcseconds. The left-hand panel is the total brightness
    temperature with the peak at about $2\times 10^{13}$\,K, with the
    arrow indicating the scintillation velocity direction. The
    right-hand panel is the polarized brightness with maximum 70\% of
    the total brightness.}\label{fig-0405pol}
\end{center}
\end{figure}

Like many other IDV sources, PKS~0405$-$385 shows more rapid and
larger fractional variations in polarization than in total intensity,
which can be interpreted as being due to two or more differently
polarized, scintillating sub-components within the total intensity
scintillating component. A detailed correlation analysis of the Stokes
I, Q, and U fluctuations of PKS~0405$-$385 observed in 1996 was
performed by \citet{ric2002}.  It was shown that the observed
fluctuations were consistent with a local enhancement in scattering at
a distance in the range of 3--70\,pc from the Earth (taking into
account uncertainty in the scintillation velocity), which is much
closer than the screen distance assumed by \citet{ked97}.  
The favoured model of \citeauthor{ric2002} placed
the screen at a distance of about 25\,pc.  The observations at 8.6 GHz
are then well modelled by scintillation of a $30 \times 22\,\mu$as
source, with about $180^{\circ}$ rotation of the polarization angle
along its long dimension as illustrated in Fig.~\ref{fig-0405pol}, from
\citet{ric2002}.  At least 3 differently polarized components are
required to fit the observations and the resulting model is not
uniquely constrained. For the model which is illustrated, the authors
chose to minimise the implied source brightness temperature, and
allowed a maximum of 70\% fractional polarization, which is close to
the theoretical maximum from a uniform synchrotron source. The model
peak brightness temperature is $2 \times 10^{13}$\,K, lower than that
inferred in the 1997 paper because of the reduced distance to the
scattering screen.

\subsection{J1819+3845}
After PKS~0405$-$385, the next IHV quasar to be discovered was the
fainter quasar J1819+3845, for which the variability was discovered
serendipitously with the WSRT \citep{dtdb2000}. With variations of up
to 10\% per minute at 5\,GHz, J1819+3845  exhibits the most dramatic
radio variability observed in an extragalactic source. If such
variations were intrinsic then the source, at $z=0.54$, would have an
angular size of order 10 nano-arcseconds and would therefore {\em
have} to scintillate, such that ISS would in any case dominate the
observed variability. Unlike PKS 0405$-$385, J1819+3845 does not
exhibit outbursts of variability but rather has continued to show its
characteristic rapid variability since discovery. The rapidity of the
variability in J1819+3845 was immediately explained as being due to an
unusually nearby scattering screen, within a few tens of pc from the
Sun \citep{dtdb2000}.

\begin{figure}[h]
\begin{center}
  \includegraphics[width=13cm]{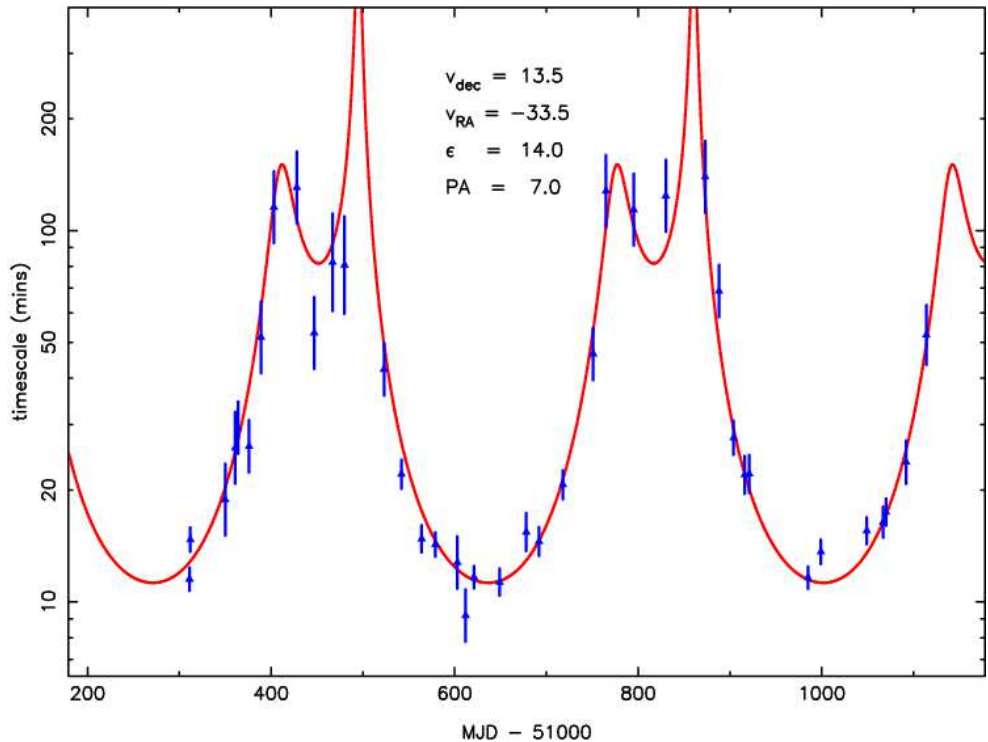}
  \caption{The large annual modulation in characteristic timescale
    observed in J1819+3845 -- note that the y-axis is shown on a
    logarithmic scale. The line shows the best fit to these
    measurements and two-station time delay data, fitting for the
    screen velocity and an elongated scintillation pattern
    \citep[from][]{dtdb2003}.}\label{fig-j1819ac}
\end{center}
\end{figure}

The discovery of a changing time delay of up to $\sim 100$ seconds
between the VLA and WSRT \citep{dtdb2002}, and of a dramatic annual
cycle in the characteristic timescale, $T_{\rm char}$, of its variations
\citep{dtdb2003}, left no doubt that the principal mechanism
responsible for the dramatic variability of J1819+3845 is also ISS.
The annual cycle results from the change in scintillation velocity due
to the Earth's orbital motion, and depends on both the transverse
velocity of the scattering plasma and the two-dimensional structure of
the scintillation pattern. An analysis of the J1819+3845 annual cycle
from more than two years' of monitoring with WSRT reveals a highly
anisotropic scintillation pattern with an axial ratio $> 6$:1
\citep{dtdb2003}. Fig.~\ref{fig-j1819ac} shows $T_{\rm char}$
measurements over two years for J1819+3845, overlaid with the model
annual cycle for the best fit screen velocity and anisotropic pattern.
From the scintillation characteristics the scattering
plasma is found to reside in a strong, thin scatterer within $\sim
10$ parsecs, which leads to a source size at 5\,GHz of 100 to
900\,$\mu$as and brightness temperature of $10^{10}$ to $10^{12}$\,K.

While the brightness temperature inferred from the annual cycle of
J1819+3845 is not especially high, there is some evidence for very
high-brightness components in the source from an analysis of the
variations at 1.4\,GHz. \citet{mdb2006} reported the discovery
of rapid, frequency dependent variations which could be modelled as
diffractive interstellar scintillation (DISS). This is the only reported case
of DISS of a quasar. If the interpretation of \citeauthor{mdb2006} is
correct, the implied brightness temperature is in excess of $2\times
10^{14}$\,K. If the brightness temperature is indeed this high, then
special emission processes may be present, e.g. cyclotron maser
emission \citep{beg2005}.

\subsection{PKS~1257$-$326}
\begin{figure}[h]
\begin{center}
 \includegraphics[width=12cm]{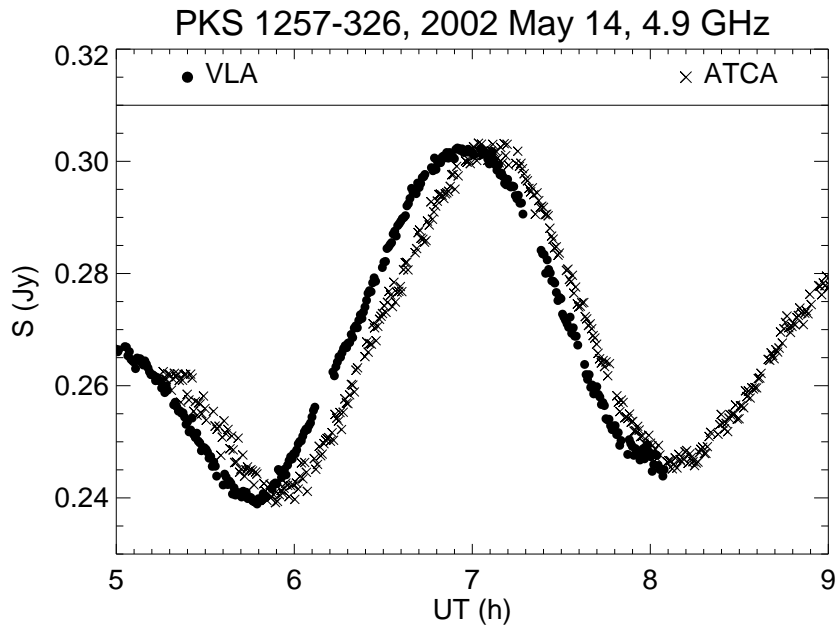}
  \caption{Simultaneous observations of PKS~1257$-$326 with the VLA
    and ATCA from 2002 May, showing a clear time delay of 8 minutes
    between the variability patterns with the VLA
    leading.}\label{fig-1257}
\end{center}
\end{figure}

The IHV quasar PKS~1257$-$326 was discovered serendipitously with the
ATCA \citep{big2003}. ATCA monitoring of PKS~1257$-$326 at 4.8 and
8.6~GHz revealed an annual cycle in the timescale of variability which
is repeated over several years of observations at both frequencies.
Successful time delay measurements have been made for PKS~1257$-$326
between the ATCA and VLA on three occasions during 2002-03
\citep{big2006}. The observed annual cycle and time delays demonstrate
conclusively that the rapid, large-amplitude variability of
PKS~1257$-$326 is entirely due to ISS. A striking feature of the time
delay measurements for PKS~1257$-$326 is the length of the delays; for
both PKS~0405$-$385 and J1819+3845 the measured delays were $\sim 2-3$
minutes at most, whereas for PKS~1257$-$326 delays as long as 8
minutes were  observed, as shown in Fig.~\ref{fig-1257}. These long
time delays, when combined with the annual cycle in variability
timescale, imply that the scintillation pattern must be highly
anisotropic, as has also been determined for the other two fast
scintillators \citep{ric2002,dtdb2003}.

In principle the annual cycle and time delay observations can be
combined to determine the peculiar velocity of the scattering medium
as well as properties of the scintillation pattern, i.e. its
characteristic length scale as well as the axial ratio and position
angle of anisotropy in the pattern. However there are degenerate
solutions when the scintillation pattern is highly anisotropic, as
shown in \citet{big2006}. The characteristic scale of the
scintillation pattern along its short axis can still be uniquely
determined, however the pattern scale and component of scintillation
velocity parallel to the long axis are degenerate.

Interestingly, PKS~1257$-$326 also showed an annual cycle in the time
offset between the 4.8 and 8.6~GHz scintillation patterns observed
over the course of 2001 \citep{big2003}. This was modelled as a
jet-like source which is optically thick between 5 and 8 GHz. That the
time offset always has the same sign implies that the scattering
screen always crosses the 8.6 GHz ``component'' first, thereby
constraining the direction of the offset. For a screen distance of
10\,pc \citep{big2006}, the fitted displacement vector corresponds to
an offset of approximately 12\,$\mu$as which at the source redshift of
$z=1.26$ corresponds to a projected linear displacement of order 0.1\,pc. More
recent monitoring data suggests the offset has changed, perhaps as a
result of  the intrinsic evolution of the source. Both the flux
density and spectral index of the source have slowly evolved over
several years of ATCA monitoring.

\section{Discussion}
In weak scattering, the characteristic length scale of the
scintillation pattern is related to the size of the first Fresnel
zone, $r_{\rm F} = \sqrt{cL/(2\pi\nu)}$, where $\nu$ is the observing
frequency and $L$ is the distance to the scattering medium which is
assumed to be confined to a plane. Thus for a given velocity of the
scattering screen, a shorter scintillation timescale, $T_{\rm char}$,
implies a closer scattering screen. When the source has angular size
larger than the Fresnel scale at the screen, however, then $T_{\rm
char}$ is increased and the amplitude modulation is reduced.  The
closer the scattering screen, the larger the angular size of the
source which can scintillate through it.  This implies that for AGN,
whose angular sizes are generally inferred to be larger than the
Fresnel scale at distances greater than a few tens of parsec, large
and rapid amplitude scintillation may {\em only} be observed through
nearby scattering screens.  The fact that only a handful of IHV
quasars has been found suggests that the covering fraction of nearby
scattering material in the Galaxy is very small. Another consideration
is that more distant scattering material can cause angular broadening
which may quench the scintillation in foreground screens.

Both source and screen properties play a role in the observed
scintillation of AGN. Of the three IHV quasars discussed here, two
show long-lived rapid scintillation over several years of monitoring,
while PKS~0405$-$385 shows episodes of IHV, lasting from months to
years. The analysis of \citet{ked2006} found no clear connection
between long-term source-intrinsic changes and episodes of IHV. This 
suggests that intermittency in ISM turbulence, rather than source
evolution, could be responsible for the episodic IDV observed in
PKS~0405$-$385.

All three IHV sources show evidence for highly anisotropic scattering
at frequencies of 5\,GHz and above. Highly anisotropic ISM turbulence
is suggested also from other observations, e.g. the parabolic arcs
observed in the secondary spectra of pulsars undergoing diffractive
scintillation \citep{wal2004}. Unfortunately, high anisotropy leads to
ambiguity in solving for the scintillation parameters, which to some
extent limits the information on source structure obtainable from ISS
observations. Nevertheless, ISS is a useful probe of source structure
on scales smaller than can be resolved with current VLBI, and also of
properties of turbulence in the local ISM.

\begin{acknowledgements}
HB thanks the organisers of ``Scattering and Scintillation in
Radioastronomy'' for the invitation to attend and present a talk at
the conference in Pushchino. We thank our colleagues, especially
Barney Rickett, Ger de Bruyn, Jane Dennett-Thorpe and Mark Walker for
sharing data and ideas presented in this paper. The ATCA is part of
the Australia Telescope, which is funded by the Commonwealth of
Australia for operation as a National Facility managed by CSIRO. The
Westerbork Synthesis Radio Telescope is operated by ASTRON
(Netherlands Foundation for Research in Astronomy) with support from
the Netherlands Foundation for Scientific Research (NWO). The National
Radio Astronomy Observatory is a facility of the National Science
Foundation operated under cooperative agreement by Associated
Universities, Inc.
\end{acknowledgements}

\end{document}